\documentclass[12pt]{article}

\def\be{\begin{equation}}
\def\ee{\end{equation}}

\usepackage{graphicx,amsmath,amssymb}
\usepackage{multirow}
\usepackage{subfigure}

\usepackage{amsmath}
\usepackage{amssymb}
\usepackage{cite}

\begin{document}
\titlepage
\begin{flushright}
IPPP/16/67 \\
\today \\
\end{flushright}

\vspace*{0.5cm}
\begin{center}
{\Large \bf  Photon--initiated  processes at high mass}

\vspace*{1cm}

L. A. Harland-Lang$^a$, V. A. Khoze$^{b,c}$ and M. G. Ryskin$^c$

\vspace*{0.5cm}
$^a$ Department of Physics and Astronomy, University College London, WC1E 6BT, UK \\           
$^b$ Institute for Particle Physics Phenomenology, Durham University, DH1 3LE, UK    \\
$^c$
 Petersburg Nuclear Physics Institute, NRC Kurchatov Institute, 
Gatchina, St.~Petersburg, 188300, Russia \\
\end{center}

\begin{abstract}
\noindent We consider the influence of photon--initiated processes on high--mass particle production. We discuss in detail the photon PDF at relatively high parton $x$, relevant to such processes, and evaluate its uncertainties. In particular we show that, as the dominant contribution to the input photon distribution is due to coherent photon emission, at phenomenologically relevant scales the photon PDF is already well determined in this region, with the corresponding uncertainties under good control.  We then demonstrate the implications of this result for the example processes of high--mass lepton and $W$ boson pair production at the LHC and FCC. While for the former process the photon--initiated contribution is expected to be small, in the latter case we find that it is potentially significant, in particular at larger masses.
\end{abstract}
\vspace*{0.5cm}

\section{Introduction}\label{sec:intro}

As we enter the era of precision LHC phenomenology, where NNLO QCD calculations are becoming the standard for many processes, the influence of electroweak corrections is increasingly relevant. A complete treatment of these inevitably requires the inclusion of diagrams with initial--state photons, with corresponding photon parton distribution function (PDF) introduced in analogy to the more commonly considered PDFs of the quarks and gluons~\cite{Gluck:2002fi,Martin:2004dh,Ball:2013hta,Martin:2014nqa,Schmidt:2015zda}. As discussed recently in~\cite{Pagani:2016caq,Bourilkov:2016qum,Accomando:2016tah,Mangano:2016jyj} the photon--initiated contribution may be significant for the production of lepton, $W$ boson and top quark pairs at higher invariant masses, and hence higher parton $x$. Such processes are of much phenomenological interest, being particularly sensitive to electroweak corrections and the PDFs, as well BSM physics; high mass lepton pair production, for example, is an irreducible background to the Drell--Yan production of a new $Z'$ boson.

The issue of how to extract the photon PDF, and what the uncertainties associated with this are, is therefore crucial to any complete discussion of these processes. A range of approaches to this problem have been taken by the global PDF fitting groups: the first attempt in the MRST2004QED set~\cite{Martin:2004dh} fixed the functional form of the photon PDF by taking a simple model for photon emission from the valence quarks, while CT14QED~\cite{Schmidt:2015zda} generalized this to allow freedom in the overall normalization, which can then be extracted from data. An alternative approach is taken in NNPDF2.3QED~\cite{Ball:2013hta} (and more recently NNPDF3.0QED~\cite{Bertone:2016ume}), where instead the photon is parameterised freely, as in the case of the quarks and gluons, and fitted to a selection of inclusive data. In this case, the corresponding PDF uncertainties due to the quite unconstraining data considered in these fits are very large.

However, these previous approaches in fact omit an important physical distinction between the photon and the quarks and gluons. The crucial difference is that QED, in contrast to QCD, corresponds to a long range force that does not suffer from the issue of non--perturbativity at low scales. Thus, a proton may coherently emit a photon ($p\to p\gamma$): such a process is experimentally extremely well measured, being governed by the well known electric and magnetic proton form factors for coherent photon emission, and is expected to constitute the dominant component of the input photon PDF. This is accounted for in the approach of~\cite{Martin:2014nqa,Harland-Lang:2016apc} (see also~\cite{Harland-Lang:2016lhw,Harland-Lang:2016qjy}), where it is shown that the photon PDF is then determined to a relatively high degree of accuracy. It is important to emphasise that the inclusion of this effect is not a theoretical assumption: indeed, exclusive lepton and $W$ boson pair production, due to precisely this initial--state coherent photon emission have been observed by both ATLAS~\cite{Aad:2015bwa,Aaboud:2016dkv} and CMS~\cite{Chatrchyan:2011ci,Chatrchyan:2013akv,Khachatryan:2016mud} at the LHC. Such processes contribute by definition to the corresponding inclusive observables.

It is therefore important to consider the consequences of these physical considerations, and the approach which derives from it for describing the photon PDF; we will consider for concreteness in this paper the cases of high mass lepton and $W$ boson pair production. Here, the $\sim \alpha^2$ suppression in the initial--state $\gamma\gamma$ luminosity may be overcome by the enhancement of the $t$--channel photon--induced process at higher masses. However, for certain sets the most significant effect is due to the PDFs themselves: the NNPDF2.3QED~\cite{Ball:2013hta,Bertone:2016ume} set in particular predicts a sharper decrease in the quark (and gluon) densities compared to the central photon value, albeit within sizeable PDF uncertainties in the latter case. In such a situation, for both processes the photon--initiated contribution is found to be potentially sizeable and even dominant at high invariant mass~\cite{Bourilkov:2016qum,Accomando:2016tah,Mangano:2016jyj}. The case of $t\overline{t}$ production has also recently been discussed in~\cite{Pagani:2016caq}, where it is again shown that the NNPDF2.3QED set is consistent with a sizeable photon--initiated contribution at larger invariance masses, as well as forward rapidities. Although for the sake of brevity we will not deal with this explicitly here, our results can be readily extended to such a process.

Given these findings, it is natural to consider what the prediction is for these processes within the approach of~\cite{Martin:2014nqa,Harland-Lang:2016apc} (see also~\cite{Gluck:2002fi}). In other words, what are the consequences of this dominantly coherent photon input PDF for  the size and uncertainties of the photon PDF at higher $x$, and what are the implications for these high mass production processes? This is the question we consider in this paper: we will show that this approach in fact predicts that the photon PDF is well constrained by the dominantly coherent input requirement, with at higher $x$ any unknown incoherent contribution, already expected to be small, being further kinematically suppressed. We will demonstrate that this follows from the relatively simple nature of the photon DGLAP evolution, which due to the small size of the coupling $\alpha$ may to very good approximation be solved exactly~\cite{Harland-Lang:2016apc,Harland-Lang:2016lhw}. Within this approach, we will find that the decrease in the $\gamma\gamma$ luminosity is qualitatively similar to the quark and gluon cases. We will also discuss how the recent ATLAS measurement of high mass lepton pair production~\cite{Aad:2016zzw} and the corresponding extraction of the photon PDF presented in this analysis lends qualitative support to our results.

The production of lepton and $W$ boson pairs are processes of much phenomenological interest at both the LHC and at a Future Circular Collider (FCC). With this in mind, we will consider cross section predictions for both of these processes at 13 and 100 TeV. In the case of lepton pair production we will show that the DY contribution is expected to be dominant out to very high masses, with a relatively small  contribution from the photon---initiated process ($\sim 10$\% for our choice of cuts). For $W$ pair production, on the other hand, the relative contribution from the photon--initiated process is more significant, and at higher masses it is comparable in size to the standard QCD--initiated process. In this case a very careful consideration of the uncertainties associated with the photon PDF is therefore essential.

Finally, during the last stages of preparing this manuscript the paper~\cite{Manohar:2016nzj} appeared. While the overall approach to treating the photon PDF and the details of the analysis are quite distinct, this work also includes the dominantly coherent input component. We may therefore expect the general conclusions to be consistent with our findings. Although we will not provide a detailed investigation of this question here, we present a brief comparison to the results of this approach. As we shall see, the predicted photon PDF does indeed quite closely coincide with our results.  Thus, we will expect comparable results to hold for the cases of high mass lepton and $W$ boson pair production when using the \texttt{LUXqed} set.

The outline of this paper is as follows. In Section~\ref{sec:photongen} we describe our approach to modelling the photon PDF, demonstrating in Section~\ref{sec:photondglap} how the DGLAP equation for the photon PDF may be solved, and then discussing in Section~\ref{sec:input} how the input photon PDF may be described. In Section~\ref{sec:photonhighx} we discuss the implications for the photon, in particular at higher $x$, and compare our results with the other available PDFs, concentrating on the NNPDFQED3.0 set. In Section~\ref{sec:dy} we present predictions for high mass lepton and $W$ boson pair production at the LHC and $\sqrt{s}=100$ TeV FCC. Finally, in Section~\ref{sec:conc} we conclude.

\section{The photon PDF: general considerations}\label{sec:photongen}

\subsection{Solving the DGLAP equation}\label{sec:photondglap}

The starting point for any discussion of the photon PDF is the corresponding DGLAP evolution equation for the distribution $\gamma(x,Q^2)$. At LO in $\alpha$ and $\alpha_S$ this is given by\footnote{In this section we work for simplicity at LO in $\alpha_S$, but this discussion can readily be generalised to NLO, as in~\cite{Harland-Lang:2016qjy}, using the results of~\cite{deFlorian:2015ujt} for the corresponding splitting functions; in the following sections we use the full NLO result.}
\be\label{dglap}
\frac{\partial \gamma(x,Q^2)}{\partial \ln Q^2}=\frac{\alpha(Q^2)}{2\pi}\!\int_x^1\!\frac{dz}z\! \left(P_{\gamma\gamma}(z)\gamma(\frac xz,Q^2)
+\sum_q e^2_qP_{\gamma q}(z)q(\frac xz,Q^2)\right)\;.
\ee
Here $P_{\gamma q}(z)$ is the $q\to \gamma$ splitting function, and $P_{\gamma\gamma}$ corresponds to the virtual self--energy correction to the photon propagator, given by
\be\label{pgg}
P_{\gamma\gamma}(z)=-\frac{2}{3}\left[N_c\sum_q e^2_q +\sum_l e^2_l\right]\delta(1-z)\;,
\ee
where $q$ and $l$ denote the active quark and lepton flavours in the fermion loop.

As the virtual correction (\ref{pgg}) is proportional to an overall delta function the corresponding contribution to (\ref{dglap}) is proportional to the photon PDF evaluated at $x$. Therefore, if we ignore the small effect that the photon PDF has on the evolution of the quark and gluons (as discussed in~\cite{Harland-Lang:2016apc}, these generally give less than a 0.1\% correction to the photon), which enter at higher orders in $\alpha$, then (\ref{dglap}) can be solved exactly, giving~\cite{Harland-Lang:2016apc,Harland-Lang:2016lhw}
\begin{align} \nonumber
\gamma(x,\mu_F^2)&=\gamma(x,Q_0^2)\,S_{\gamma}(Q_0^2,\mu_F^2)+\int_{Q_0^2}^{\mu_F^2}\frac{\alpha(Q^2)}{2\pi}\frac{dQ^2}
{Q^2}\int_x^1\frac{dz}z \;\sum_q e^2_qP_{\gamma q}(z)q(\frac xz,Q^2)\,S_{\gamma}(Q^2,\mu_F^2)\;,\\  \label{gampdf}
&\equiv \gamma^{{\rm in}}(x,\mu^2)+\gamma^{\rm evol}(x,\mu^2)\;,
\end{align}
where $\gamma(x,Q_0^2)$ is the input PDF at the scale $Q_0$, and we have introduced the photon Sudakov factor
\be\label{sudgam}
S_{\gamma}(Q_0^2,\mu_F^2)=\exp\left(-\frac{1}{2}\int_{Q_0^2}^{\mu_F^2}\frac{{\rm d}Q^2}{Q^2}\frac{\alpha(Q^2)}{2\pi}\int_0^1 {\rm d} z\sum_{a}\,P_{a\gamma}(z)\right)\;.
\ee
Here $P_{q(l)\gamma}(z)$ is the $\gamma$ to quark (lepton) splitting function, given by
\be
P_{a\gamma}(z)=N_a\left[z^2+(1-z)^2\right]\;,
\ee
where $N_a=N_c e_q^2$ for quarks and $N_a=e_l^2$ for leptons, while the factor of $1/2$ in (\ref{sudgam}) is present to avoid double counting over the quark/anti--quarks (lepton/anti--leptons). Written in this form, the physical interpretation of the Sudakov factor is clear: it represents the Poissonian probability for no parton emission from the photon during its evolution from the low scale $Q_0$ up to the hard scale $\mu_F$. 

Thus, the photon PDF (\ref{gampdf}) at $\mu_F$ can be written as the sum of a contribution from low--scale emission of a photon, with no further branching, and a term due to higher scale DGLAP emission from quarks\footnote{This separation was used in~\cite{Harland-Lang:2016apc} to demonstrate how a rapidity gap veto can be included in photon--initiated processes.}. For the purposes of the discussion in this paper, the crucial point is that when considering the photon PDF and its corresponding uncertainty at some given $x$ and $\mu_F^2$ value, the contributions to this from the input photon distribution, at the starting scale $Q_0$, and from the DGLAP evolution term due to high scale emission from the quarks, are completely separated; this will greatly simplify the discussion which follows, and allow some fairly simple and robust conclusions to be drawn.

\subsection{The input distribution}\label{sec:input}

The photon PDF has been separated in (\ref{gampdf}) into an input component at $Q_0$ and an evolution component, due to high scale $q\to q\gamma$ emission. While the latter quantity is given in terms of the generally well determined quark PDFs, the former quantity is on the face of it completely unknown. Thus the uncertainty on the photon distribution at some scale  $\mu_F$ is given quite directly in terms of the uncertainty on the starting distribution $\gamma^{\rm in}$, and it is this object which we are principally interested in.

 It is perfectly possible to simply treat this as an unknown quantity in a global fit, i.e. in the same way as the quarks and gluons. This is the approach taken in the latest NNPDFQED fit~\cite{Ball:2013hta,Bertone:2016ume}, where the freely parameterised photon is fitted to DIS and a small set of LHC data, namely $W,Z$ and high/low--mass Drell--Yan production (more precisely this is achieved by Bayesian reweighting, see~\cite{Ball:2013hta} for full details). Due to the generally small contribution from photon--initiated processes, the constraining power of this data is quite limited, and the corresponding PDF uncertainties are large. 

However, by treating the photon PDF identically to the quark and gluons, a significant part of the available experimental information is in fact being thrown away~\cite{Gluck:2002fi,Martin:2014nqa,Harland-Lang:2016apc}. The crucial difference is that QED, in contrast to QCD, corresponds to a long range force that does not suffer from the issue of non--perturbativity at low scales. Thus a proton may coherently emit a photon ($p\to p\gamma$) at low scale $Q<Q_0\sim $ 1 GeV, and this will contribute directly to the input component in (\ref{gampdf}). Such a process is experimentally extremely well measured, being governed by the well known electric and magnetic proton form factors for coherent photon emission. In particular we have
\begin{equation}\label{gamcoh}
\gamma_{\rm coh}(x,Q_0^2)=\frac{1}{x}\frac{\alpha}{\pi}\int_0^{Q^2<Q_0^2}\!\!\frac{{\rm d}q_t^2 }{q_t^2+x^2 m_p^2}\left(\frac{q_t^2}{q_t^2+x^2 m_p^2}(1-x)F_E(Q^2)+\frac{x^2}{2}F_M(Q^2)\right)\;,
\end{equation}
where  $q_t$ is the transverse momentum of the emitted photon, and $Q^2$ is the modulus of the photon virtuality, given by
\begin{equation}\label{qmincoh}
Q^2=\frac{q_t^2+x^2m_p^2}{1-x}\;,
\end{equation}
The functions $F_E$ and $F_M$ are
\begin{equation}\label{form1}
F_M(Q^2)=G_M^2(Q^2)\;,\qquad F_E(Q^2)=\frac{4m_p^2 G_E^2(Q^2)+Q^2 G_M^2(Q^2)}{4m_p^2+Q^2}\;,
\end{equation}
with
\begin{equation}\label{form2}
G_E^2(Q^2)=\frac{G_M^2(Q^2)}{7.78}=\frac{1}{\left(1+Q^2/0.71 {\rm GeV}^2\right)^4}\;,
\end{equation}
in the dipole approximation, where $G_E$ and $G_M$ are the `Sachs' form factors, which have been very precisely measured in a range of elastic $ep$ scattering experiments\footnote{The dipole approximation describes the available data to within a few percent in the lower $Q^2$ region most relevant to our study, however this description is not perfect, and a completely precise calculation would go beyond this and in addition should consider the uncertainties associated with the available form factor data. For the purposes of this paper, however, such a high level of precision is not necessary.}. As the contribution to $ep$ scattering for low photon $Q^2$ is dominantly coherent, we expect (\ref{gamcoh}) to give the dominant contribution to the input photon PDF. Thus the input photon distribution is in fact already well determined.

More precisely, in general there will also be some contribution from incoherent emission ($\gamma p \to \gamma X$), where the proton dissociates after the scattering process. That is, we have
\begin{equation}\label{inputdef}
\gamma(x,Q_0^2)=\gamma_{\rm coh}(x,Q_0^2)+\gamma_{\rm incoh}(x,Q_0^2)\;,
\end{equation}
where the second term corresponds to this incoherent input; it is this combined input PDF, including both coherent and incoherent components, which corresponds to the freely parameterised NNPDF distribution described above. In general, as recently discussed in~\cite{Manohar:2016nzj} this incoherent contribution may be constrained from experimental data on $F_2$ and $F_L$, however for our considerations it is sufficient to use a simplified model which gives an upper bound on such a contribution. Thus, following~\cite{Gluck:2002fi,Martin:2014nqa} we model this emission process as being due to one photon emission from the valence quarks in the leading--logarithmic approximation; such an approach is also taken in~\cite{Martin:2004dh,Schmidt:2015zda} to model the photon PDFs, although in these cases no coherent component is included. We take\footnote{In fact, we take the slightly different form described in footnote 3 of~\cite{Martin:2014nqa}, with as in (\ref{gamincoh}) the replacement $F_1(Q^2)\to G_E(Q^2)$ made to give a more precise evaluation for the probability of coherent emission.}
\begin{equation}\label{gamincoh}
\gamma_{\rm incoh}(x,Q_0^2)=\frac{\alpha}{2\pi}\int_x^1\frac{{\rm d}z}{z}\left[\frac{4}{9}u_0\left(\frac{x}{z}\right)+\frac{1}{9}d_0\left(\frac{x}{z}\right)\right]\frac{1+(1-z)^2}{z}\int^{Q_0^2}_{Q^2_{\rm min}}\frac{{\rm d}Q^2}{Q^2+m_q^2}\left(1-G_E^2(Q^2)\right)\;,
\end{equation}
where
\begin{equation}\label{qminincoh}
Q^2_{\rm min}=\frac{x}{1-x}\left(m_\Delta^2-(1-x)m_p^2\right)\;,
\end{equation}
accounts for the fact that the lowest proton excitation is the $\Delta$--isobar, and the final factor $(1-G_E^2(Q^2))$ corresponds to the probability to have no intact proton in the final state (which is already included in the coherent component). Here $m_q=m_d$($m_u$) when convoluted with $d_0$($u_0$), and the current quark masses are taken. 
Crucially, as the quark distributions are frozen for $Q<Q_0$, this represents an upper bound on the incoherent contribution. If we consider the momentum fraction
\be 
p_\gamma = \int {\rm d}x\, x\gamma(x,Q_0^2)\;,
\ee
carried by the photon at the starting scale $Q_0^2=2\,{\rm GeV}^2$, then even for this upper bound we find
\be\label{pgam}
p_\gamma^{\rm coh} =0.15\% \quad\quad p_\gamma^{\rm incoh.}=0.05\%\;, 
\ee
that is we expect $p_\gamma^{\rm incoh.} \ll p_\gamma^{\rm coh.}$, consistent with the general expectation that the emission process for low $Q^2$ photons should be dominantly coherent. As the coherent input is quite precisely determined, the uncertainty on the input photon PDF is essentially purely due to the incoherent term. Being maximally conservative, we can consider a range of incoherent inputs, with the lower bound being simply setting $\gamma^{\rm incoh.}(x,Q_0^2)=0$, and the upper bound calculated as described above. Taking this as our uncertainty band, we then expect from (\ref{pgam}) a $\sim \pm 10-15\%$ uncertainty on the photon PDF at the starting scale $Q_0$. On the other hand, at higher scales as the contribution from the evolution term in (\ref{gampdf}) becomes more significant, this uncertainty will be smaller; we will show this explicitly in the following section, where we will see that for phenomenologically relevant scales the uncertainty due to the incoherent input shrinks to $\sim \pm 5\%$. It should be emphasised that this is a relatively conservative estimate of the uncertainty on the photon PDF due to the incoherent input component. In particular, it is possible and desirable to further constrain this incoherent input in a global fit, while a more complete treatment accounting for example for the ($\Delta$...) resonant contribution to the incoherent input, and more generally constraining this with the existing $ep$ scattering data will also further reduce this uncertainty. 

Nonetheless, even within this simplified and conservative approach, we can see that the corresponding uncertainty on the photon PDF is already under relatively good control. In contrast, the NNPDF3.0QED set~\cite{Bertone:2016ume} gives
\be
p_{\gamma}=(1.26\pm 1.26)\%\;,
\ee
that is, a $\sim 100\%$ uncertainty, with a central value which lies much higher than that expected from the simple physical considerations above; we will see the impact of this is in the following sections.

\section{The photon PDF: results}\label{sec:photonhighx}

\begin{figure} 
\begin{center}
\includegraphics[scale=0.75]{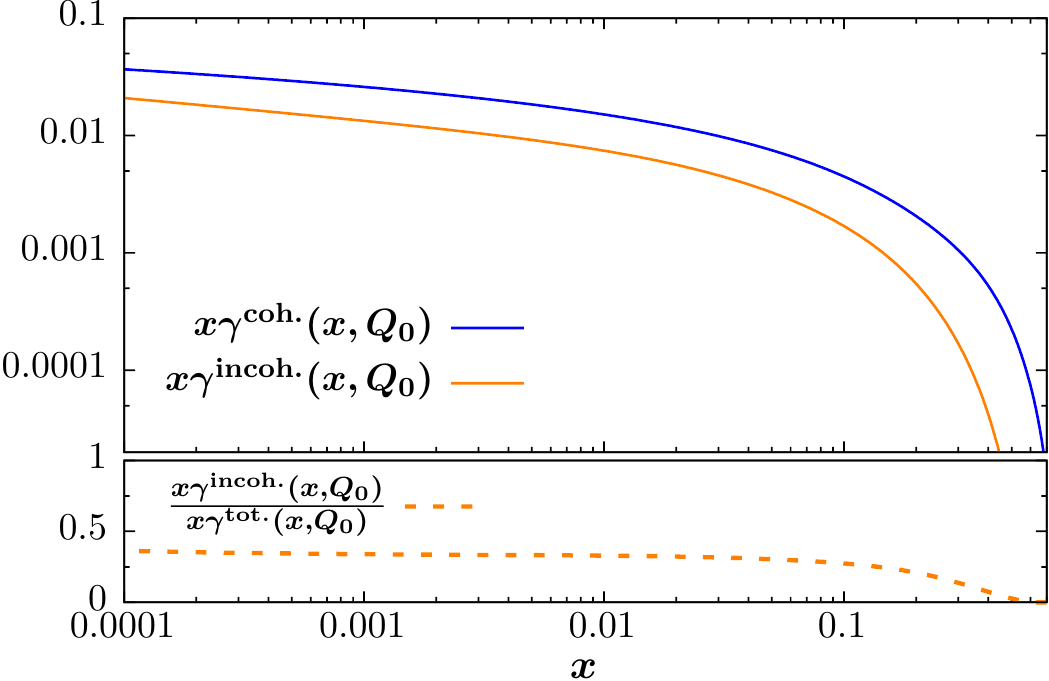}\quad
\includegraphics[scale=0.66]{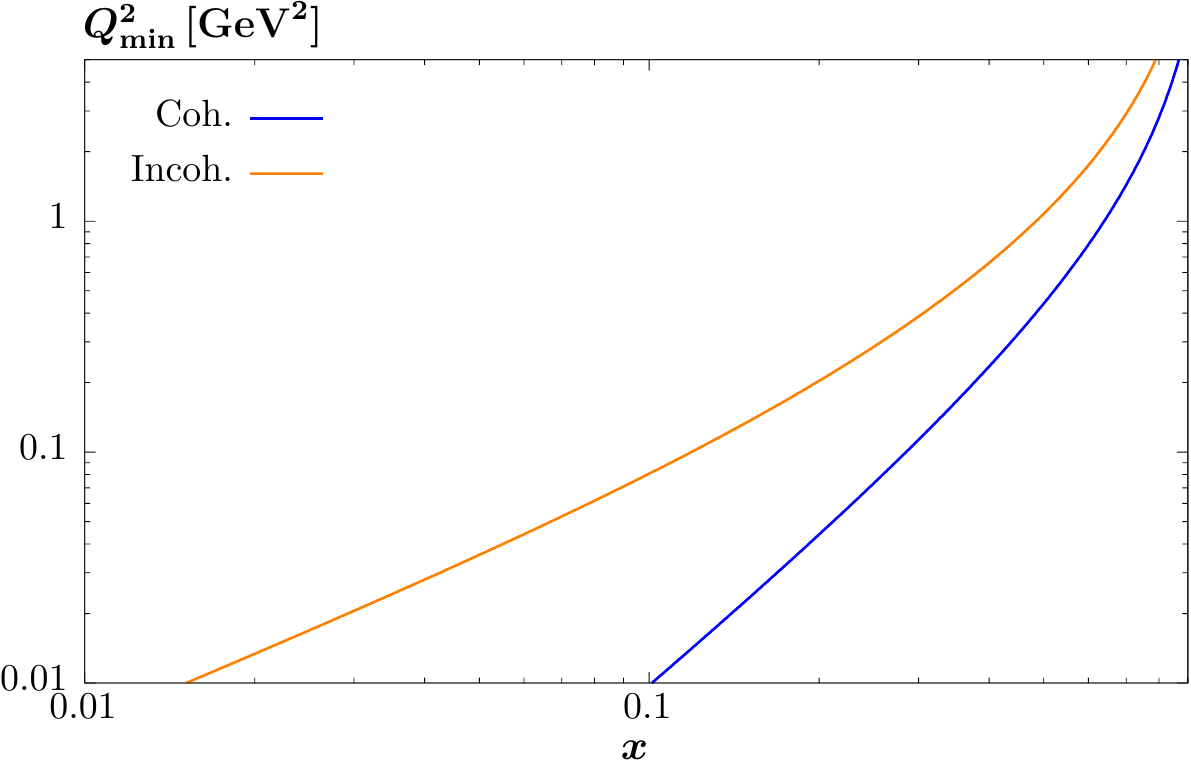}
\caption{(Left) The coherent (\ref{gamcoh}) and incoherent (\ref{gamincoh}) components of the photon PDF at the starting scale $Q_0^2=2\, {\rm GeV}^2$. (Right) The minimum photon $Q^2$ vs. the momentum fraction $x$ carried for the coherent (\ref{gamcoh}) and incoherent (\ref{gamincoh}) emission processes.}
\label{fig:gam2q}
\end{center}
\end{figure}

In the previous section we demonstrated how the photon PDF at the starting scale is already quite precisely determined in terms of the form factors for  coherent $p\to p \gamma$ emission. We will now demonstrate the impact this result has 
on the photon PDF at higher $x$ values. We show in Fig.~\ref{fig:gam2q} (left) the contributions from the coherent (\ref{gamcoh}) and incoherent (\ref{gamincoh}) components of the photon PDF at the starting scale $Q_0^2=2\, {\rm GeV}^2$, where as described above the incoherent term corresponds to an upper bound on this contribution. Here, and in all results which follow, we take treat the evolution (\ref{dglap}) at NLO in $\alpha_s$. We make use of the MMHT2014NLO~\cite{Harland-Lang:2014zoa} set for the quark PDFs in the incoherent component, as well as for the PDFs in the evolution of the photon.

As expected from the previous discussion, the incoherent component is smaller than the well determined coherent component, and constitutes $\sim 25\%$ of the total photon PDF at intermediate values of $x$, consistent with (\ref{pgam}). However, interestingly the ratio of incoherent to coherent is found to decrease with increasing $x$, such that in the higher $x\gtrsim 0.1$ region, the coherent component is particularly dominant.  This is due in part to the decreasing phase space for photon emission from the individual quarks, with the range of the $z$ integral in (\ref{gamincoh}) decreasing with increasing $x$. In addition to this, another physical effect is playing a role, due to the minimum photon virtuality $Q^2_{\rm min}$, given by (\ref{qmincoh}) and (\ref{qminincoh}). The kinematic minimum (\ref{qmincoh}) follows simply from the on--shellness requirement for the outgoing proton, and similarly in (\ref{qminincoh}) for the outgoing $\Delta$ resonance. In the latter case this corresponds to the mass of the lowest lying resonance above the proton: for higher mass resonance production, the kinematic minimum will be larger still. The effect of this is shown in Fig.~\ref{fig:gam2q} (right). Due to the higher mass of the dissociating state in the latter incoherent case, the minimum photon virtuality can be quite large for higher $x$, with the effect that contribution from low--scale incoherent photon emission becomes kinematically limited; by construction we must have $Q^2<Q_0^2\sim 1\,{\rm GeV^2}$, while the contribution for photon $Q^2>Q_0^2$ is given by the evolution component in (\ref{gampdf}), in terms of the relatively well constrained quark PDFs. This effect can be seen in Fig.~\ref{fig:gam2q} (left) in the high $x$ region, where the coherent component becomes completely dominant.

\begin{figure} 
\begin{center}
\includegraphics[scale=0.85]{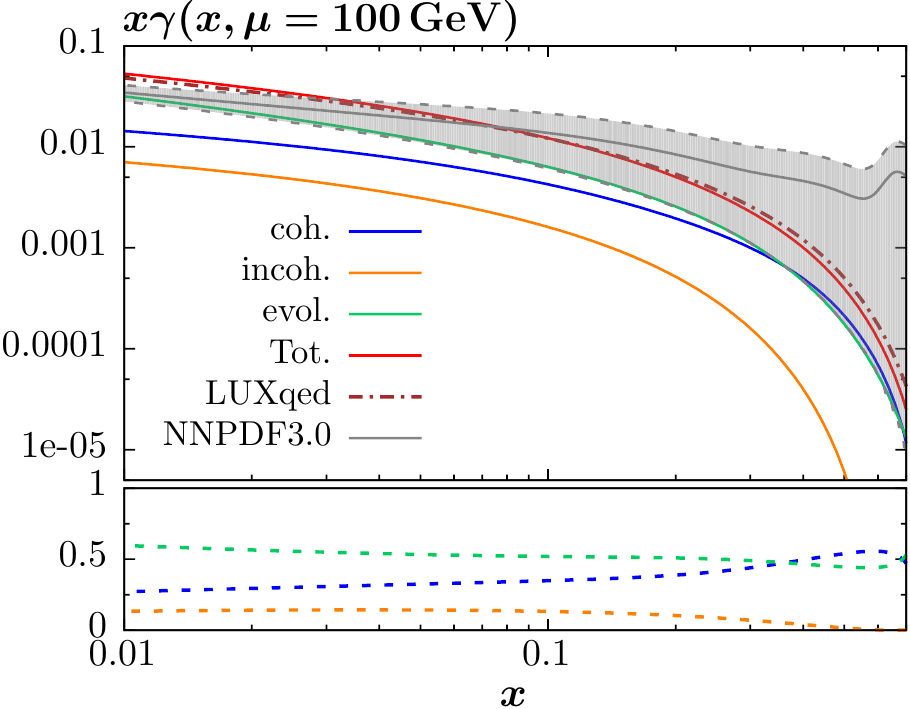}
\includegraphics[scale=0.85]{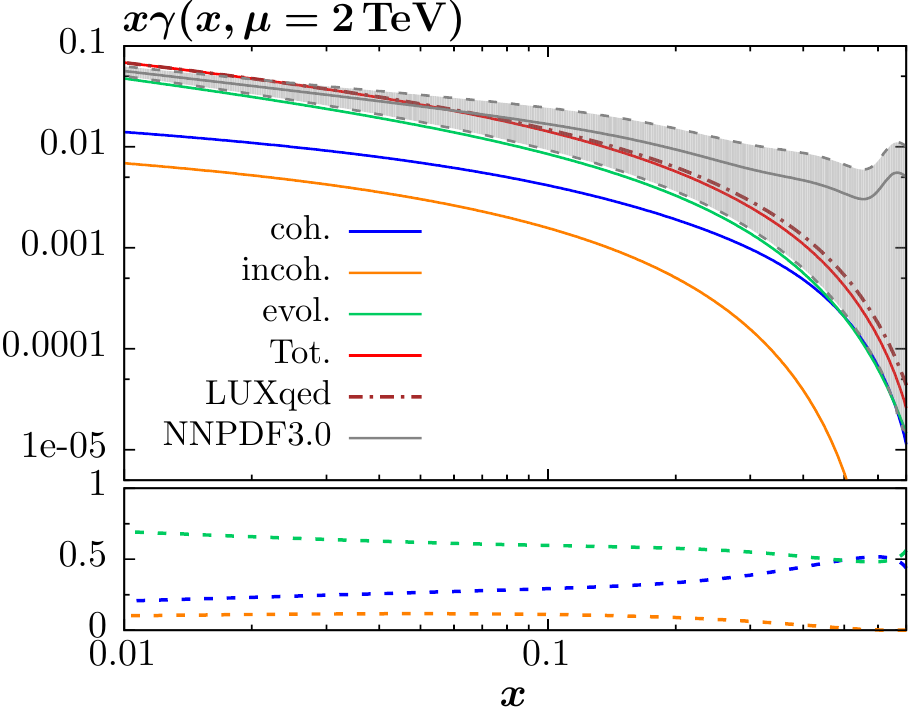}
\caption{The photon PDF at scale $\mu_F=100$ and 2000 GeV, with the breakdown between coherent, incoherent and evolution components, defined as in (\ref{gampdf}) and (\ref{inputdef}) given. Also shown is the NNPDF3.0~\cite{Bertone:2016ume} result, with the corresponding $68\%$ C.L. uncertainty bands, and the LUXqed~\cite{Manohar:2016nzj} prediction. In the lower plots the ratios of the different components to the total photon PDF are shown.}
\label{fig:pdfcomp}
\end{center}
\end{figure}

In Fig.~\ref{fig:pdfcomp} (right) we show the photon PDF at $\mu_F=2$ TeV, corresponding to $x\sim 0.2$ at the LHC, with the contributions from the input coherent and incoherent, and evolution components shown explicitly. We also show in  Fig.~\ref{fig:pdfcomp} (left) the corresponding PDF for the lower scale  $\mu_F=10$ GeV, to give an indication of the influence of the photon evolution on these different components. We can see that the effect of evolution is as expected to further decrease the contribution from the incoherent input, which is already $\lesssim 10\%$ of the total photon at $\mu_F=100$ GeV, and even lower for $\mu_F=2$ TeV. Thus in the cross section results which follow we expect a conservative $\sim \pm 5\%$ uncertainty due to this effect. The coherent contribution, even at the higher scale $\mu_F=2$ TeV is $\sim 20\%$ at $x\sim 0.01$, and increases to $\sim 50\%$ at higher $x$; for  $\mu_F=100$ GeV it is larger still. The \texttt{LUXqed} prediction is shown and is found to coincide quite closely with our result, although some deviation is visible, in particular at higher $x$.

The NNPDF3.0 distribution, with the corresponding $68\%$ C.L. uncertainty bands, is also shown in Fig.~\ref{fig:pdfcomp}: here, and in all results which follow, we take the NLO set with $\alpha_s(M_Z)=0.118$. For the lower $x$ region the contribution from the high scale $q\to q\gamma$ evolution component in (\ref{gampdf}) is dominant, and as a result the corresponding uncertainties are under reasonable control\footnote{The slight deviation between our results and the NNPDF sets, even accounting for the PDF uncertainties, at lower $x$ is due to the differing `truncated' solution to the DGLAP equation applied in the latter case.}. As $x$ increases, however, the phase space for the DGLAP $q\to q\gamma$ emission process decreases, and the contribution from the coherent photon input becomes more important.  This effect is evident in the NNPDF set, where the increasing contribution from the poorly determined input photon leads to a rapidly increasing uncertainty as $x$ increases.

\begin{figure} 
\begin{center}
\includegraphics[scale=0.65]{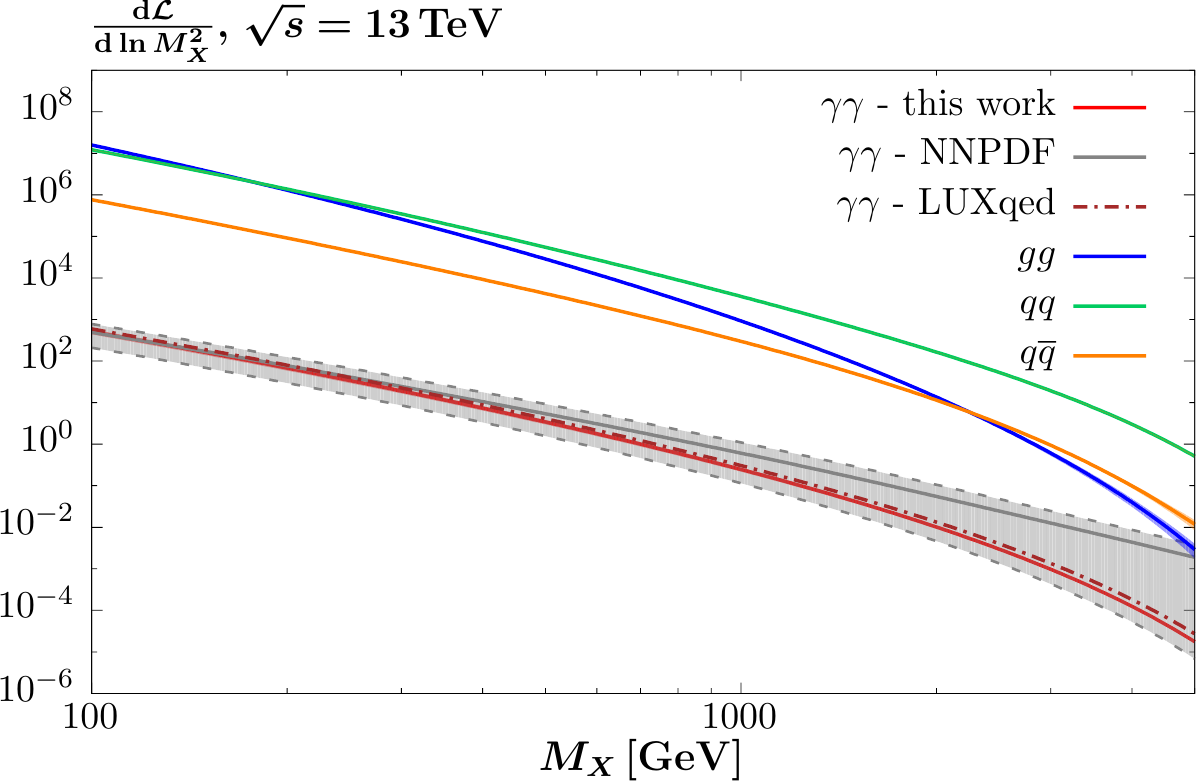}
\includegraphics[scale=0.65]{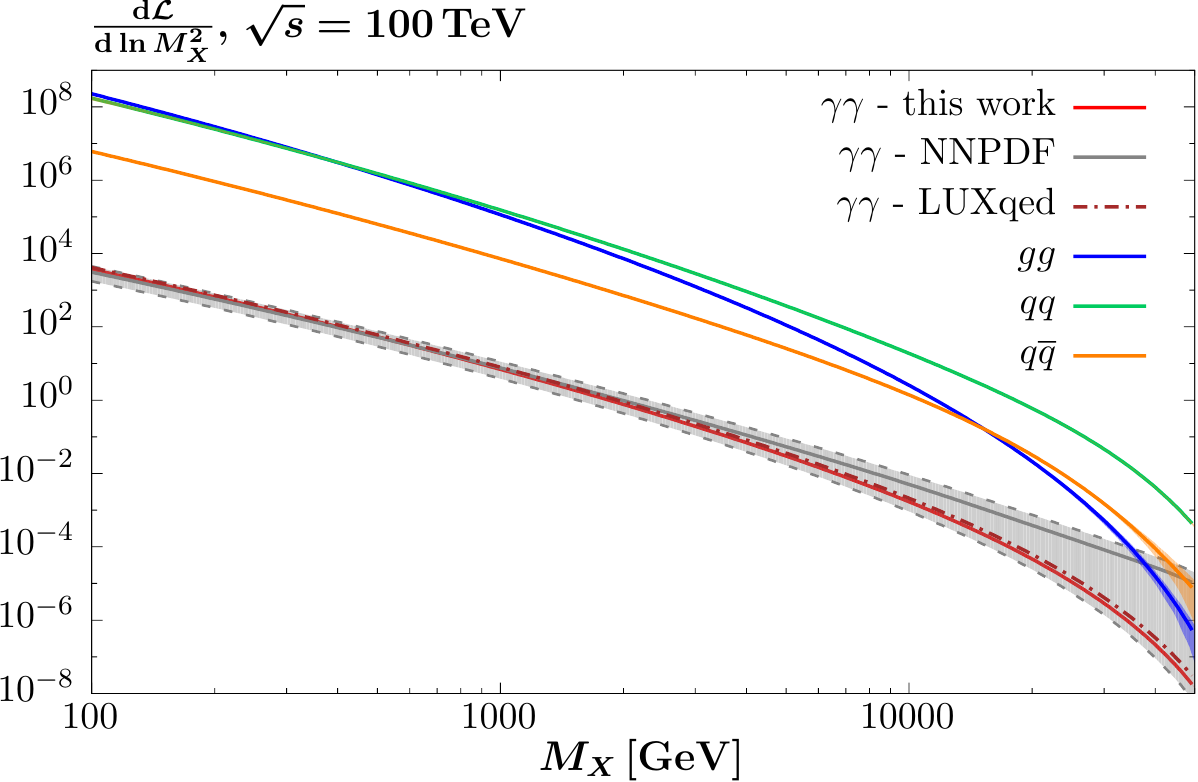}
\caption{The  $\gamma\gamma$, $gg$, $q\overline{q}$ and $qq$ PDF luminosities. The $\gamma\gamma$ case is shown for the NNPDF3.0~\cite{Bertone:2016ume} set and following the approach of Section~\ref{sec:input} , while all other luminosities correspond to the NNPDF set. The corresponding $68\%$ C.L. uncertainty bands are shown in the NNPDF cases, while an uncertainty band due to varying the incoherent component between $x\gamma(x,Q_0)=0$ and the upper bound of (\ref{gamincoh}) is shown, although barely visible, for our prediction. The LUXqed~\cite{Manohar:2016nzj} prediction is also shown.}
\label{fig:pdflumi}
\end{center}
\end{figure}

In Fig.~\ref{fig:pdflumi} we show the corresponding PDF luminosities, defined as
\begin{equation}\label{lumi}
\frac{{\rm d}\mathcal{L}_{ij}}{{\rm d} \ln M_X^2}= \frac{M_X^2}{s}\int^1_{\tau} \frac{{\rm d}x_1}{x_1} \,f_i(x_1,M_X^2)f_j(\tau/x_1,M_X^2)\;,
\end{equation}
where $\tau=M_X^2/s$ and $f_i$ is the corresponding PDF for parton $i$. As well as the $\gamma\gamma$ case discussed above, we also show for comparison the $qq$, $q\overline{q}$ (defined in both cases as a uniform sum over the 5 corresponding quark flavours) and $gg$ cases, using the same NNPDF set. For our prediction, we now for illustration include an uncertainty band due to varying the incoherent component between $x\gamma(x,Q_0)=0$ and the upper bound of (\ref{gamincoh}), although in the plots this is essentially invisible within the width of the central curves. Other uncertainties, due for example to the quark (and at higher orders, gluon) PDFs entering the photon evolution in (\ref{dglap}), the use of the dipole approximation (\ref{form2}) for the elastic form factor and the choice of $Q_0$ in (\ref{gampdf}) are not included here. These effects are expected to be generally subleading in comparison to that due to the incoherent input, and will be omitted in the results which follow. Nonetheless, it is worth bearing in mind that the effect of these will be to increase the total uncertainty on the photon PDF somewhat, which should be accounted for in a complete analysis; for the current purposes, however, this is not necessary. The \texttt{LUXqed} prediction is shown and is again found to coincide quite closely with our result, with some deviation at higher $M_X$.

The same increase in Fig.~\ref{fig:pdflumi} in the NNPDF uncertainty band at high $M_X$ for the $\gamma\gamma$ case is clear. However, interestingly we can see that the trend in the central value of the NNPDF $\gamma\gamma$ luminosity is remarkably different compared to the other partons, with the former decreasing much less rapidly at high $M_X$, i.e. high $x$. On the other hand, our prediction shows no such significant difference, and roughy follows the same trend as in the quarks. As discussed in~\cite{Mangano:2016jyj} some steepening of the PDF luminosities for the QCD partons may be expected due to the differing behaviours of $\alpha_{\rm QED}$ and $\alpha_s$ at higher scales. However this effect, which is indeed observable in particular upon comparison of our result for the $\gamma\gamma$ and the $gg$ luminosity, is relatively small and cannot explain the difference seen in the NNPDF case. We are therefore led to conclude that this potentially significant difference is an artefact of the large uncertainties in the NNPDF photon PDF; the physically motivated photon PDF of our approach, which lies towards the lower end of the NNPDF uncertainty band, displays no significant difference in behaviour at higher $x$ compared to the quarks and gluons.

It is therefore in this higher $x$ region that the importance of including all available information about the photon PDF is clearest; by excluding the additional input which comes from considering the physics of the dominantly coherent photon emission process at the starting scale $Q_0$, the corresponding PDF uncertainties are dramatically over--inflated. By including this information, as in Section~\ref{sec:input}, the predicted photon PDF at higher $x$ is determined quite precisely to lie close to the lower edge of the NNPDF uncertainty band. It has for example been pointed out in~\cite{Bourilkov:2016qum,Accomando:2016tah} that the upper limits on the NNPDF photon PDF predict potentially sizeable photon--initiated contributions to the Drell--Yan cross section at high mass, with large corresponding PDF uncertainties. From the above considerations, however, we expect this not to be the case: we will consider this in more detail in the following section.

\section{Cross section predictions}\label{sec:dy}

As discussed in the introduction, the photon--initiated contribution to lepton and $W$ boson pair production may be particularly significant at higher mass, where the production cross sections are relatively enhanced due to the $t$--channel nature of the corresponding Feynman diagrams. We therefore consider predictions for both these processes at the LHC and FCC in this section. We use our own implementation of these processes, with the corresponding LO cross sections given by (see e.g.~\cite{Brodsky:1970vk,Tupper:1980bw})
\begin{align}\nonumber
\frac{{\rm d}\sigma}{{\rm d}\cos\theta^*}(\gamma\gamma\to W^+W^-) &= \frac{\pi\alpha^2\beta}{\hat{s}}\frac{19-6\beta^2(1-\beta^2)+2(8-3\beta^2)\beta^2\cos^2\theta^*+3\beta^4\cos^4\theta^*}{(1-\beta^2\cos^2\theta^*)^2} \,, \\
\frac{{\rm d}\sigma}{{\rm d}\cos\theta^*}(\gamma\gamma\to l^+l^-) &= \frac{2\pi\alpha^2\beta}{\hat{s}}\frac{1+2\beta^2(1-\beta^2)(1-\cos^2\theta^*)-\beta^4\cos^4\theta^*}{(1-\beta^2\cos^2\theta^*)^2}\,,
\end{align}
where $\beta=(1-4m^2/\hat{s})^{1/2}$, with $m=m_{W},m_{l}$, and $\theta^*$ is the angle of the outgoing particles with respect to the photons in the $\gamma\gamma$ C.M. frame.  As we are interested in the high mass regime, in the lepton case we will for concreteness take the massless limit, $m_l=0$, in what follows.

\begin{figure} 
\begin{center}
\includegraphics[scale=0.65]{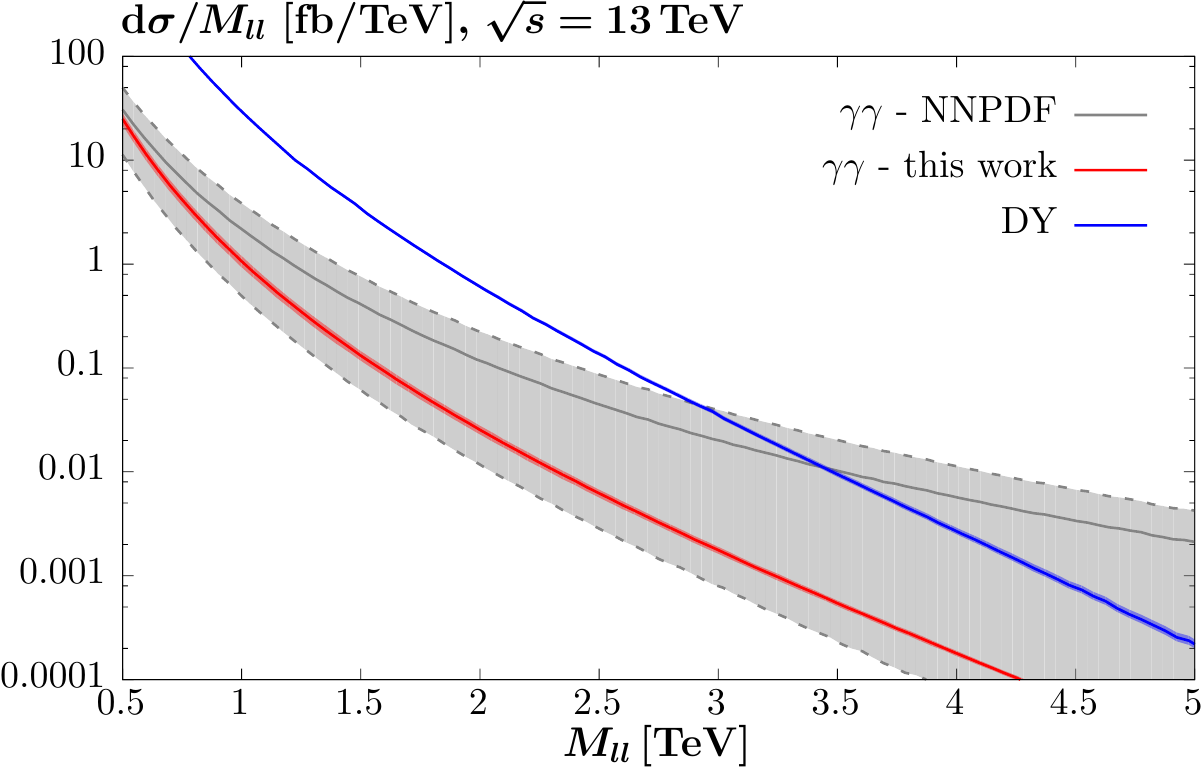}
\includegraphics[scale=0.65]{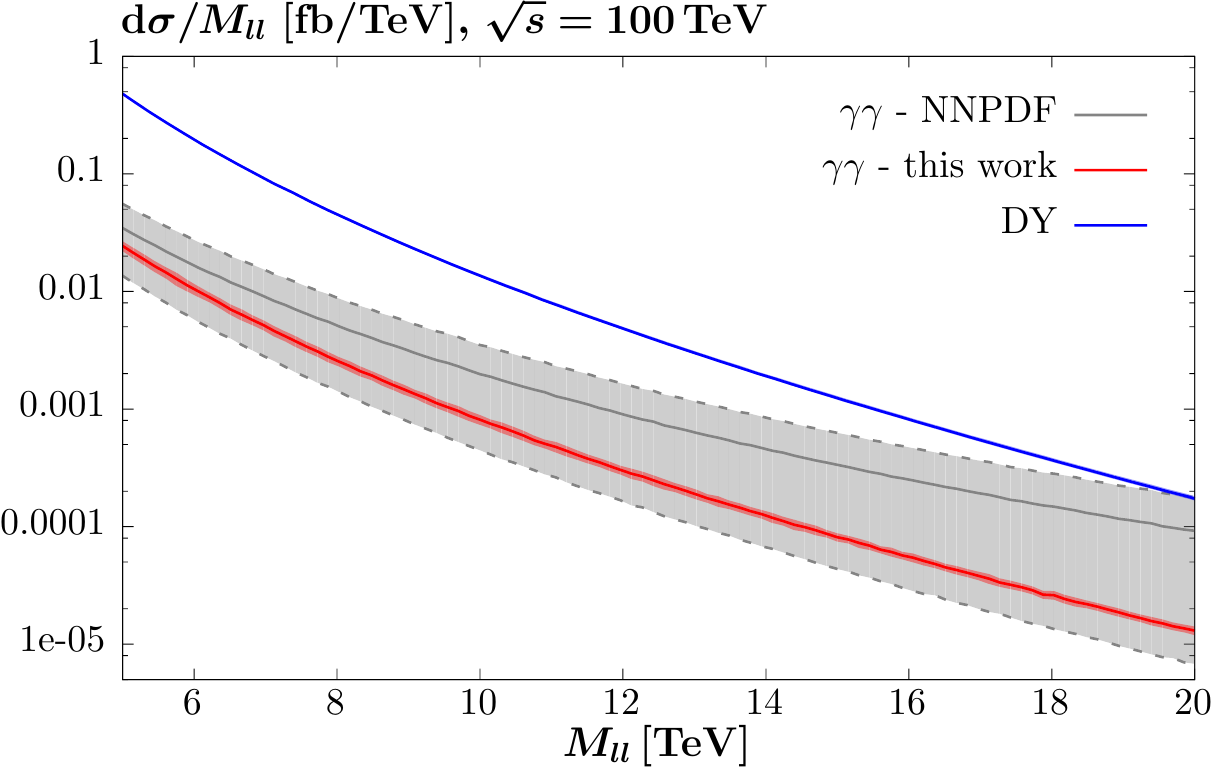}
\caption{The differential lepton pair production cross sections at $\sqrt{s}=13$ TeV and 100 TeV with respect to the invariant mass of the pair $M_{ll}$, for lepton $|\eta|<2.5$ and $p_\perp> 20$ GeV. The photon--initiated contributions predicted following the approach of Section~\ref{sec:input} and the NNPDF3.0QED~\cite{Bertone:2016ume} set, including the 68\% C.L. uncertainty bands are shown, in addition to the NLO Drell--Yan cross section, calculated with \texttt{MCFM}~\cite{Campbell:2011bn}. An uncertainty band due to varying the incoherent component between $x\gamma(x,Q_0)=0$ and the upper bound of (\ref{gamincoh}) is shown for our prediction.}
\label{fig:cross13}
\end{center}
\end{figure}

In Fig.~\ref{fig:cross13} we show the lepton pair production cross section via the photon--initiated production and Drell--Yan production mechanisms, for lepton $|\eta|<2.5$ and $p_\perp> 20$ GeV. The former is shown using both the approach of Section~\ref{sec:input} and with the NNPDF3.0QED set, while the latter is calculated at NLO in $\alpha_s$ with  \texttt{MCFM}~\cite{Campbell:2011bn} using MMHT2014NLO~\cite{Harland-Lang:2014zoa} PDFs; the results which follow are not affected significantly by NNLO corrections. For the curve corresponding to our approach, we show an uncertainty band due to varying the incoherent component between $x\gamma(x,Q_0)=0$ and the upper bound of (\ref{gamincoh}). The 68\% PDF uncertainty bands are shown for the \texttt{MCFM} predictions in all cases which follow, although for certain distributions this is sufficiently small that it is not visible on the plots.

For the LHC predictions in Fig.~\ref{fig:cross13} (left), we can see that for higher $M_{ll}$ the photon--initiated cross section predicted by the NNPDF set may be comparable in size and even larger than the Drell--Yan cross section, within the increasingly large PDF uncertainty bands. This was recently discussed in~\cite{Accomando:2016tah}, where it was pointed out that the potential dominance of the photon--initiated NNPDF prediction induced a large uncertainty in the predicted cross section for high mass lepton pair production; this could, for example, have an impact on searches for new heavy particles decaying to lepton pairs. However, it is our finding that this is not the case. In particular, we can see from Fig.~\ref{fig:cross13} that even up to the highest $M_{ll}$ values the predicted contribution from the photon--initiated process is fairly small, $\sim 10\%$ of the Drell--Yan. This result is entirely consistent with the expectations from Fig.~\ref{fig:pdfcomp}. Thus we expect no significant contamination from the photon--initiated process. For the FCC case shown in Fig.~\ref{fig:cross13} (right), which was recently discussed in~\cite{Mangano:2016jyj}, a similar trend is seen. Moreover, it is worth emphasising that for both the LHC and FCC cases, tighter cuts on the lepton transverse momentum $p_\perp$ and pseudorapidity $\eta$ will further decrease the relative contribution from the photon--initiated process, which being due to the $t$ and $u$ channel diagrams is more strongly peaked towards low $p_\perp$ and high $\eta$.

\begin{figure} 
\begin{center}
\includegraphics[scale=0.65]{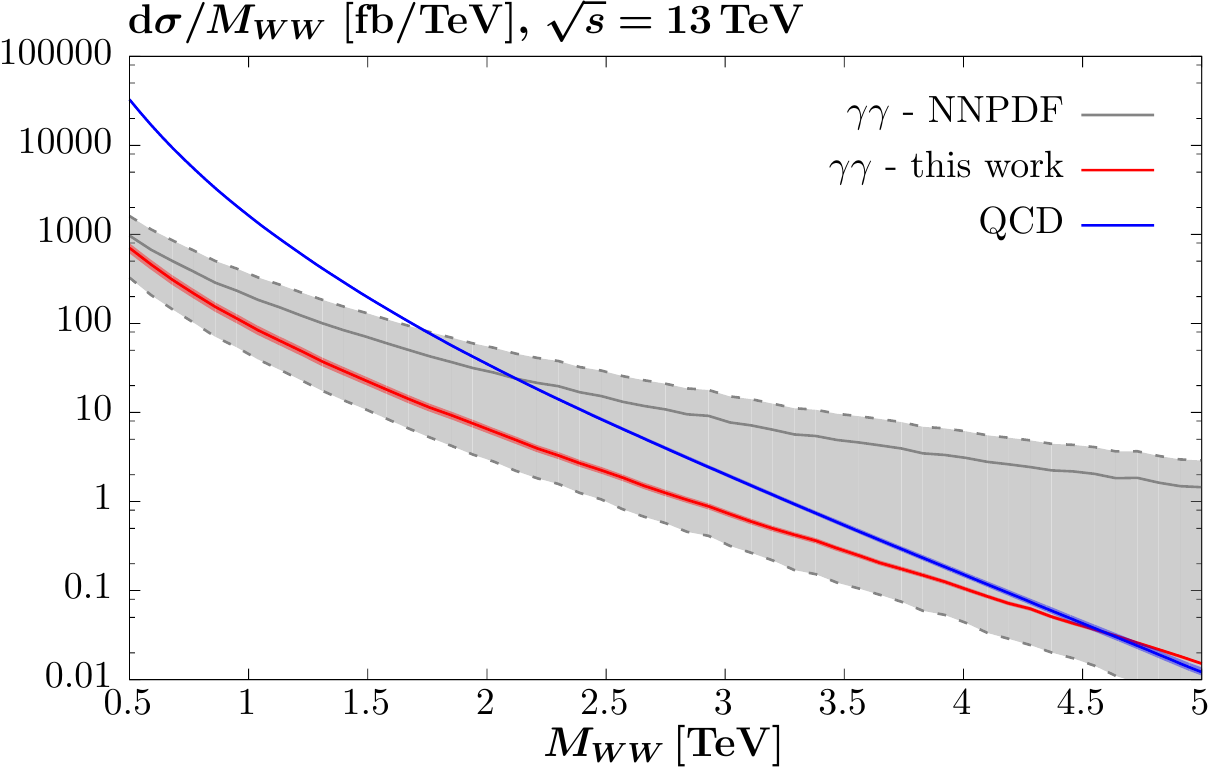}
\includegraphics[scale=0.65]{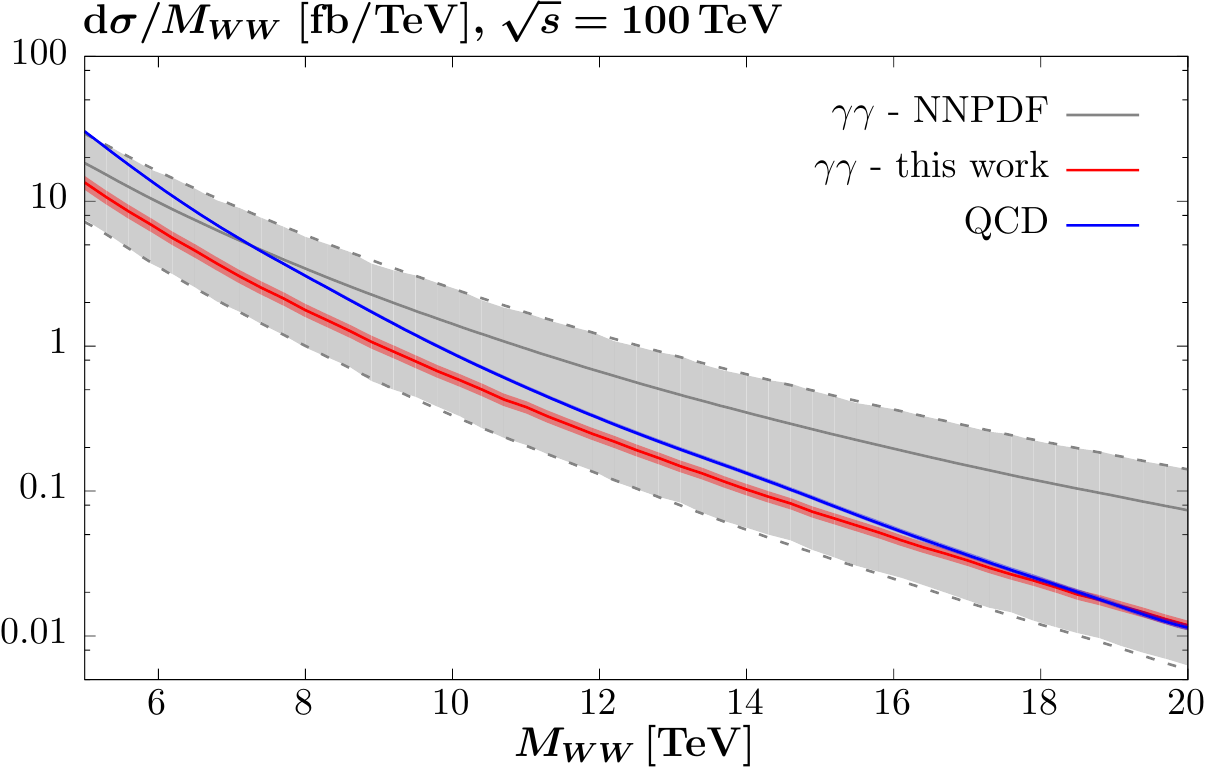}
\caption{The differential $W$ boson pair production cross sections at $\sqrt{s}=13$ TeV and 100 TeV with respect to the invariant mass of the pair $M_{WW}$, for $W$ pseudorapidity $|\eta|<4$. The photon--initiated contributions predicted following the approach of Section~\ref{sec:input} and the NNPDF3.0QED~\cite{Bertone:2016ume}, including the 68\% C.L. uncertainty bands are shown, in addition to the NLO QCD cross section, calculated with \texttt{MCFM}~\cite{Campbell:2011bn}, and including the gluon--initiated box contribution. An uncertainty band due to varying the incoherent component between $x\gamma(x,Q_0)=0$ and the upper bound of (\ref{gamincoh}) is shown for our prediction.}
\label{fig:crossw}
\end{center}
\end{figure}

In Fig.~\ref{fig:crossw} we show predictions for the $W$ boson pair production cross sections, again at the LHC and FCC. We impose the same cuts on the $W$ boson pseudorapidities, and include no further decays, as in~\cite{Mangano:2016jyj}, for the sake of comparison. \texttt{MCFM}~\cite{Campbell:2011bn} with MMHT2014NLO~\cite{Harland-Lang:2014zoa} PDFs is used to generate the the QCD $WW$ production process at NLO in $\alpha_s$, with the $gg$--initiated box contribution also included. Again a similar trend is clear, with the NNPDF set predicting potentially a completely dominant photon--initiated contribution at higher masses, within very large uncertainties. However, for the LHC our approach predicts instead that the standard QCD--initiated is dominant, apart from at the very highest masses. On the other hand for the FCC this is no longer the case: over the mass range considered the $\gamma\gamma$ and QCD--initiated contributions are generally expected to be comparable in size. In this case a very careful treatment of the photon PDF uncertainties will be essential.

It should be emphasised that the predicted cross sections within our approach lie entirely within the NNPDF uncertainty bands, and are therefore completely consistent with these. The issue is simply that the NNPDF approach, by omitting the physical constraints on the photon input described in Section~\ref{sec:input}, allows in principle unphysically large input photon distributions, the effect of which becomes increasingly dominant at higher $x$, where the contribution from the DGLAP $q\to q\gamma$ emission from the quarks becomes smaller. On the other hand, the NNPDF starting distribution effectively parameterises the contribution from both the coherent and incoherent input components as in (\ref{inputdef}), but without any further constraints, and therefore we fully expect consistency within both approaches, once the PDF uncertainties have been properly included; it is encouraging to find that this is indeed the case. Moreover, as further data from the LHC is included in the NNPDF fit, we fully expect this consistency to continue as the PDF uncertainties decrease. 

\begin{figure} 
\begin{center}
\includegraphics[scale=0.65]{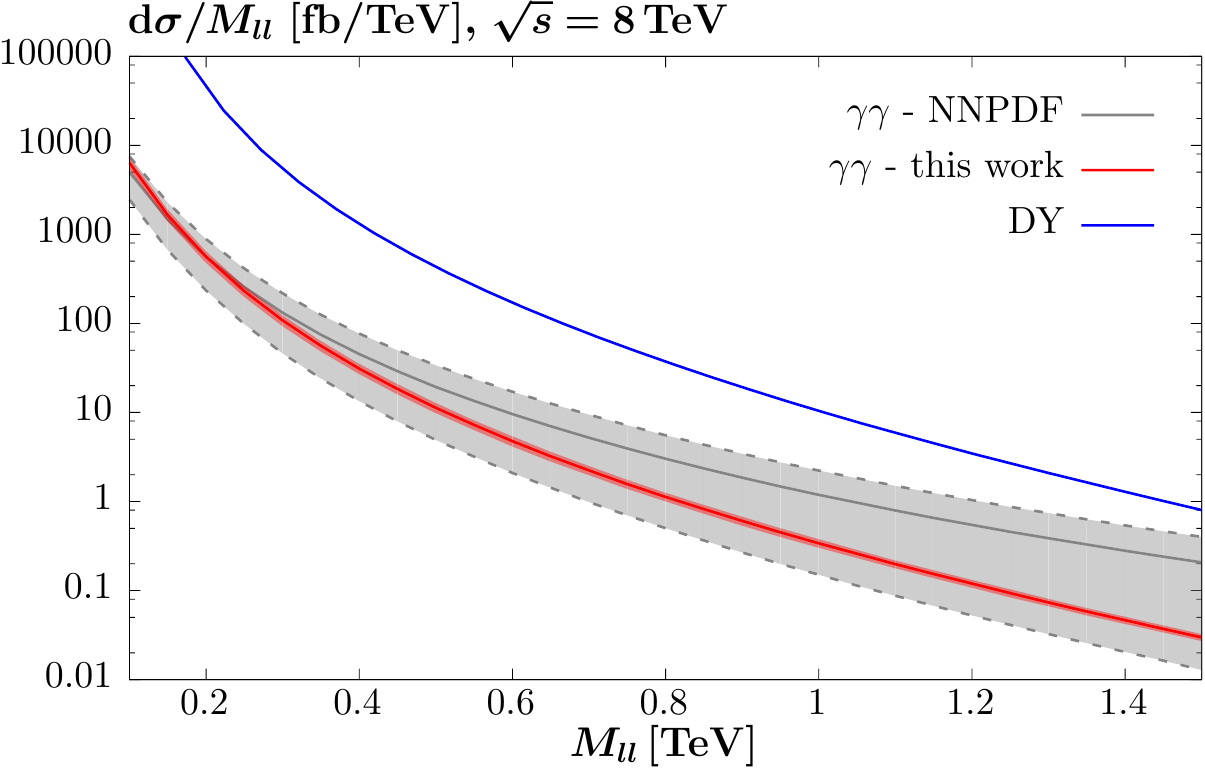}
\caption{The differential lepton pair production cross section at $\sqrt{s}=8$ TeV with respect to the invariant mass of the pair $M_{ll}$, calculated as in Fig.~\ref{fig:cross13}.}
\label{fig:cross8}
\end{center}
\end{figure}

Indeed, this expectation is supported by the recent ATLAS measurement~\cite{Aad:2016zzw} of high mass lepton pair production in the $116<M_{ll}<1500$ GeV region at $\sqrt{s}=8$ TeV, corresponding to $0.015\lesssim x \lesssim 0.2$. We show the predicted cross section for this mass region in Fig.~\ref{fig:cross8}, and we can see that for larger $M_{ll}$ quite significant photon--initiated contributions are allowed within the NNPDF uncertainty bands. However, in~\cite{Aad:2016zzw} a Bayesian reweighting exercise is performed, and it is found that a photon PDF which lies systematically on the lower end of the NNPDF2.3 uncertainty band is preferred, with greatly reduced uncertainties, see in particular Fig.~13 of~\cite{Aad:2016zzw}\footnote{It is worth emphasising that in reality it is only for $x \gtrsim 0.05$, i.e. $M_{ll}\gtrsim 400$ GeV, that the ATLAS data begin to place any significant constraint  on the photon. Thus the form of the reweighted photon distribution below $x\sim 0.05$ should not be taken too literally.}. While a full refit would be required to make a completely firm conclusion, this nonetheless provides a strong indication that the larger photon--initiated cross section predictions allowed by the higher $x$ NNPDF photon uncertainty band are already disfavoured by this ATLAS data, while our result, which predict lower photon--initiated cross sections, are qualitatively consistent.

Further experimental support for this result at the LHC can be found in the ATLAS~\cite{Aaboud:2016dkv} and CMS~\cite{Chatrchyan:2013akv,Khachatryan:2016mud} measurements of exclusive two--photon induced $W^+W^-$ production. Here, events are selected by demanding a dilepton vertex with no additional associated charged tracks within the tracker acceptance ($\eta<$ 2.4). As part of these measurements, a larger data sample of $\mu^+\mu^-$ events is also selected with the same track veto applied. Crucially, such a veto effectively isolates the photon--initated contribution; the standard Drell--Yan production process will dominantly produce additional tracks centrally, and the remaining contribution can be suppressed with further cuts and then subtracted using MC simulation. The photon--initiated cross section for this semi--exclusive case was considered in~\cite{Harland-Lang:2016apc}, where it was demonstrated how (\ref{gampdf}) could be relatively simply modified to account for such a rapidity gap veto. Predictions for semi--exclusive $\mu^+\mu^-$ production within the ATLAS/CMS event selection were presented, and it was shown that the data are completely consistent with the approach described in this paper, but are in strong tension with the higher cross sections allowed by the NNPDF2.3 set.

\section{Conclusion}\label{sec:conc}

Processes with initial--state photons are becoming increasingly relevant for phenomenology at the LHC, in particular as the requirement for high precision becomes standard. Given this, it is becoming increasingly important to constrain as precisely as possible the corresponding photon PDF. This is particularly so for processes such as high mass lepton and $W$ boson production, where the contribution from the photon--initiated process may be significant.

In this paper we have described how the photon PDF may we be quite precisely constrained by relatively simple considerations about the nature of the initial--state photon, see~\cite{Martin:2014nqa,Harland-Lang:2016apc}. In particular, QED, in contrast to QCD, corresponds to a long range force that does not suffer from the issue of non--perturbativity at low scales, and thus a proton may coherently emit a photon ($p\to p\gamma$) at low scales. This process is already experimentally very well measured in $ep$ collisions, and is governed by the well known electric and magnetic proton form factors for coherent photon emission. Taking these as input, we have shown how the photon PDF can already be well constrained with even a quite conservative model estimation for an upper limit on the remaining incoherent contribution to the photon at the starting scale: the uncertainty due to this is found to be generally $\sim 10\%$ or less, with the precise value depending on the scale and parton $x$. A more detailed treatment of the (resonant and non--resonant) incoherent contribution, and the inclusion of further constraints within the context of a global fit will reduce this even further. Indeed, as discussed in the recent analysis of~\cite{Manohar:2016nzj}, it is possible to place quite precise constraints by using low $Q^2$ inelastic structure function data. In this work, a dominantly coherent component is also included, and we have seen that the resultant photon PDF lies quite close to our prediction.

These constraints are not included in the currently available global fits~\cite{Martin:2004dh,Ball:2013hta,Schmidt:2015zda} and may lead to a signifiant overestimation in the uncertainty in the photon PDF. This can have important phenomenological implications, and indeed this is the case for high mass lepton and $W$ boson pair production, which have recently been discussed in~\cite{Bourilkov:2016qum,Accomando:2016tah,Mangano:2016jyj}. Here, it was found that the NNPDF2.3QED~\cite{Ball:2013hta} set, which freely parameterises the input photon and fits to a selection of DIS and LHC data, is consistent with strongly dominant photon--initiated contributions to both of these processes at higher masses, at the LHC and FCC. In this paper we have shown that this is not expected to be the case. We have found that the DY contribution to lepton pair production is dominant out to very high masses, while for $W$ boson pair production the QCD process is generally larger, although for higher masses the $\gamma\gamma$--initiated contribution is comparable in size. In this latter case, therefore, a precise treatment of the photon PDF will be essential.
\newline



\section*{Acknowledgements}

We thank Valerio Bertone, Amanda Cooper--Sarkar, David D'Enterria and Juan Rojo for useful discussions. VAK thanks the Leverhulme Trust for an Emeritus Fellowship. The work of MGR  was supported by the RSCF grant 14-22-00281. LHL thanks the Science and Technology Facilities Council (STFC) for support via the grant award ST/L000377/1. MGR thanks the IPPP at the University of Durham for hospitality. 

\bibliography{references}{}
\bibliographystyle{h-physrev}

\end{document}